\documentclass[preprint,aps]{revtex4}
\usepackage{mathrsfs}

\usepackage{graphicx}

\begin{document}

\title{Coexistence of Fermi Arcs and Fermi Pockets in High Temperature Cuprate Superconductors}

\author{Jianqiao Meng$^{1}$, Guodong Liu$^{1}$, Wentao Zhang$^{1}$, Lin Zhao$^{1}$, Haiyun Liu$^{1}$,
Xiaowen Jia$^{1}$, Daixiang Mu$^{1}$, Shanyu Liu$^{1}$, Xiaoli
Dong$^{1}$, Jun Zhang$^{1}$, Wei Lu$^{1}$, Guiling Wang$^{2}$, Yong
Zhou$^{2}$, Yong Zhu$^{2}$, Xiaoyang Wang$^{2}$, Zuyan Xu$^{2}$,
Chuangtian Chen$^{2}$ and X. J. Zhou$^{1,*}$}

\affiliation{
\\$^{1}$National Laboratory for Superconductivity, Beijing National Laboratory for Condensed
Matter Physics, Institute of Physics, Chinese Academy of Sciences,
Beijing 100190, China
\\$^{2}$Technical Institute of Physics and Chemistry, Chinese Academy of Sciences, Beijing 100190, China
}
\date{May 14, 2009}
%
% The abstract goes here
%
%%\begin{abstract}

%%\end{abstract}

%%\pacs{74.72.Hs, 74.25.Jb, 79.60.-i, 71.38.-k}

\maketitle

\newpage

{\bf In the pseudogap state of the high$-$T$_c$ copper-oxide
(cuprate) superconductors\cite{TimuskReview}, angle-resolved
photoemission (ARPES) measurements have seen an Fermi arc, i.e., an
open-ended gapless section in the large Fermi
surface\cite{MarshallArc,NormanArc,KShenArc,KanigelArc,WSLee,HossainYBCO,YangPocket},
rather than a closed loop expected of an ordinary metal. This is all
the more puzzling because Fermi pockets (small closed Fermi surface
features) have been suggested from recent quantum oscillation
measurements\cite{Doiron,Bangura,Leboeuf,Yelland,Jaudet,Sebastian}.
The Fermi arcs have worried the high$-$T$_c$ community for many
years because they cannot be understood in terms of existing
theories. Theorists came up with a way out in the form of
conventional Fermi surface pockets associated with competing order,
with a back side that is for detailed reasons invisible by
photoemission\cite{ArcInterpretations}.  Here we report ARPES
measurements of La-Bi2201 that give direct evidence of the Fermi
pocket. The charge carriers in the pocket are holes and the pockets
show an unusual dependence upon doping, namely, they exist in
underdoped but not overdoped samples. A big surprise is that these
Fermi pockets appear to coexist with the Fermi arcs. This
coexistence has not been expected theoretically and the
understanding of the mysterious pseudogap state in the high-T$_c$
cuprate superconductors will rely critically on understanding such a
new finding. }

The high resolution Fermi surface mapping (Fig. 1a) on the
underdoped La-Bi2201 UD18K sample using VUV laser reveals three
Fermi surface sheets with low spectral weight (labeled as LP, LS and
LPS in Fig. 1a) in the covered momentum space, in addition to the
prominent main Fermi surface (LM). One particular Fermi surface
sheet LP crosses the main band LM, forming an enclosed loop, an
Fermi pocket, near the nodal region. Quantitative Fermi surface data
measured from both VUV laser (Fig. 2a) and Helium discharge lamp
(Fig. 2b) are summarized in Fig. 2c. All the possible umklapp
bands\cite{Umklapp}, shadow bands\cite{Shadow}, and umklapp bands of
the shadow bands that may be present in the bismuth-based compounds,
are shown for comparison (see Supplementary and SFig. 2 for more
details). It is clear that the LP and LPS bands observed in VUV
laser measurement (Fig. 2a) and HP band observed in Helium lamp
measurement (Fig. 2b) are intrinsic; they can not be attributed to
any of the umklapp bands or shadow bands (Fig. 2c). The location of
the three bands can be well connected by the same superlattice
vector, indicating that the HP and LPS bands correspond to the first
order umklapp bands of the main Fermi pocket LP. The shape and area
of the Fermi pockets are also consistent in these two independent
measurements, making a convincing case on the presence of the Fermi
pocket. We note that in both the laser (Fig. 2a) and Helium lamp
(Fig. 2b and SFig. 2 in the Supplementary) measurements, all the
observed bands except for the ``Fermi pocket bands" can be assigned
by only one regular superstructure wavevector (0.24,0.24). The
presence of additional superstructure, which would give rise to new
bands, appears to be unlikely because there is no indication of such
additional bands observed in our measurements.

The Fermi pocket is observed both in the normal state and
superconducting state, as shown in Fig. 3 for the La-Bi2201 UD18K
sample. Moreover, its location, shape and area show little change
with temperature (Figs. 3a and 3f).  Below T$_c$, the opening of
superconducting gap is clearly observed on both the main band (Fig.
3c) and the back-side of the Fermi pocket (Fig. 3e).  Above T$_c$, a
portion of the main Fermi surface near the nodal region becomes
gapless as indicated by the pink solid circles on the main Fermi
surface (Fig. 3f). This is reminiscent to the Fermi arc (an
open-ended gapless section in the large Fermi surface) formation in
previous ARPES results which show an increase in the arc length with
increasing temperature\cite{NormanArc,KanigelArc}. The back side of
the Fermi pocket also becomes gapless above T$_c$ (Fig. 3j), forming
a gapless Fermi pocket loop with part of the main band. We note that
the Fermi arc on the main band appears to extend longer than the
Fermi pocket section (Fig. 3f), giving rise to an interesting
coexistence of Fermi arc and Fermi pocket.

The Fermi pocket exhibits an unusual doping dependence, as seen in
Fig. 4.  It is observed not only in the UD18K sample (Fig. 4b), but
also in the UD26K sample (Fig. 4c). But it is not seen in the UD3K
sample (Fig. 4a) and OP32K optimally-doped sample (Fig.
4d)\cite{JQMengdWave}. This peculiar doping dependence of the Fermi
pocket, i.e., it can only be observed in a limited doping range in
the underdoped region, lends further support to its intrinsic
nature. The quantitative main Fermi surface and Fermi pocket at
various dopings are summarized in Fig. 4i. The area enclosed by the
Fermi pockets in the UD18K (blue ellipse in Fig. 4i) and UD26K (grey
ellipse in Fig. 4i) samples corresponds to a doping level of 0.11
and 0.12 holes$\slash$site, respectively, which is in good agreement
with the estimated hole concentration of each
sample\cite{JQMengCrystal,AndoBi2201}.

The Fermi pocket we have observed is hole-like. This makes it
impossible to correspond to the electron-like Fermi pocket suggested
from quantum oscillation experiments\cite{Leboeuf,ChakravartyPNAS}.
Moreover, the observed Fermi pocket is not symmetrically located in
the Brillouin zone, specifically, it is not centered around
($\pi$$\slash$2,$\pi$$\slash$2) (Fig. 4i). This is distinct from
that reported in Nd$-$LSCO system\cite{NdLSCO} which is symmetrical
with respect to the ($\pi$,0)$-$(0,$\pi$) line, similar to the
``shadow band" commonly observed in Bi2201 and Bi2212\cite{Shadow}.
This particular location makes it impossible to originate from the
{\it d}-density-wave ``hidden order" that gives a hole-like Fermi
pocket centered around ($\pi$/2,$\pi$/2)
point\cite{ChakravartyPRB,ChakravartyPNAS}. Among other possible
origins of Fermi pocket formation\cite{WenLee,FCZhang,Kaul}, the
phenomenological resonant valence bond picture\cite{FCZhang} shows a
fairly good agreement with our observations, in terms of the
location, shape and area of the hole-like Fermi pocket and its
doping dependence.  In particular, the predicted Fermi pockets are
pinned at ($\pi$/2,$\pi$/2) point\cite{FCZhang} that is consistent
with our experiment (Fig. 4i).  One obvious discrepancy is that the
spectral weight on the back side of the Fermi pocket near
($\pi$/2,$\pi$/2) is expected to be zero from these
theories\cite{WenLee,FCZhang} which is at odds with our measurements
(Figs. 3d and 3i). We note that the existence of incommensurate
density wave could also potentially explain the observed pockets.
One possible wave-vector needed is (1$\pm$0.092,1$\pm$0.092) which
is diagonal that can be examined by neutron or X-ray scattering
measurements. There are indications of the charge-density-wave (CDW)
formation reported in the Bi2201 system\cite{HudsonSTM}; whether the
Fermi surface reconstruction caused by such a CDW order or its
related spin-density-wave order can account for our observation
needs to be further explored.

One peculiar characteristic of the Fermi pocket is its coexistence
with the large underlying Fermi surface (Figs. 3 and 4).  One may
wonder whether this could originate from sample inhomogeneity, i.e,
the large Fermi surface is caused by a phase at high doping while
the Fermi pocket is from a phase at low doping. We believe this is
unlikely because there is no indication of two distinct
superconducting phases detected in the magnetization measurement
(SFig. 1 in the Supplementary). In addition, no Fermi pocket is
identified for the underdoped UD3K sample (Fig. 4a).  A more
surprising observation is the coexistence of Fermi arcs and Fermi
pockets in the normal state because it has not been expected
theoretically.  These new findings will provide key insight for
understanding the anomalous pseudogap state in high-T$_c$ cuprate
superconductors.

\vspace{3mm}

\noindent {\bf Supplementary Information}  is linked to the online
version of the paper at www.nature.com/nature.

\vspace{3mm}

\noindent {\bf Acknowledgements} We thank D.-H. Lee, P. A. Lee, S.
Sachdev, Z.-X. Shen, X. G. Wen, Z. Y. Weng, T. Xiang, G. M. Zhang
and  F. C. Zhang for helpful discussions. This work is supported by
the NSFC, the MOST of China, and Chinese Academy of Sciences.

\vspace{3mm}

\noindent {\bf Author Contributions} J.Q.M. contributed to La-Bi2201
sample growth with the assistance of G.D.L.. X.L.D. and W.L.
contributed to the magnetic measurement of samples. G.D.L, W.T.Z,
L.Z., H.Y.L., J.Q.M., X.L.D., X.W.J., D.X.M., S.Y.L, J.Z., G.L.W.,
Y.Zhou, Y.Zhu, X.Y.W., Z.Y.X. and C.T.C contributed to the
development and maintenance of Laser-ARPES system. J.Q.M. carried
out the experiment with the assistance from G.D.L., W.T.Z., L.Z.,
H.Y.L., X.W.J., D.X.M., S.Y.L.. X.J.Z. and J.Q.M. analysed the data
and wrote the paper. X.J.Z. are responsible for overall project
direction, planning, management and infrastructure.

\vspace{3mm}

\noindent {\bf Author Information} Reprints and permissions
information is available at www.nature.com/reprints. Correspondence
and requests for materials should be addressed to X.J.Z.
(XJZhou@aphy.iphy.ac.cn).

\newpage

\begin{figure}[tbp]
\begin{center}
\includegraphics[width=1.0\columnwidth,angle=0]{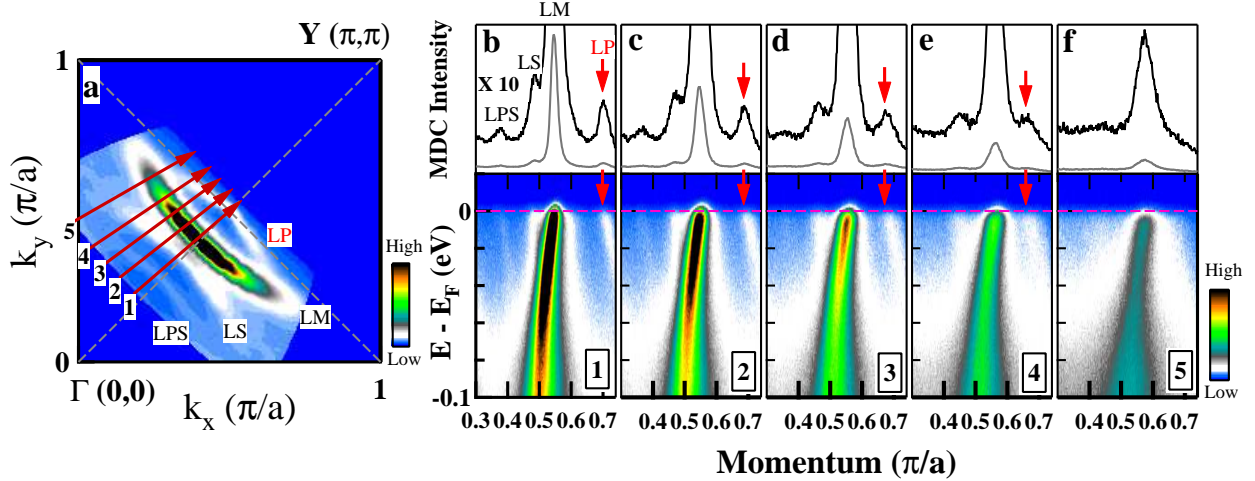}
\end{center}
\caption{ Fermi surface and band structure of a La-Bi2201 sample. In
the present paper, we use the phrase ``Fermi surface" loosely to
denote the momentum space locus of high intensity low-energy
spectral weight. In cuprate superconductors, the Fermi surface is
usually measured in the superconducting state in spite of the gap
opening because the sharpness of the features at low temperature
facilitates the precise determination of the underlying Fermi
surface as compared to the normal state. It has been shown that the
``underlying Fermi surface" determined from the minimum gap locus in
the superconducting state is identical to that in the normal
state\cite{HDingFS}. (a). Photoemission intensity at the Fermi
energy (E$_F$) as a function of k$_x$ and k$_y$ for the La-Bi2201
UD18K sample (underdoped, T$_c$=18 K) measured at a temperature of
14 K. It is obtained by symmetrizing the original data with respect
to the (0,0)-($\pi$,$\pi$) line. Four Fermi surface sheets are
resolved in the covered momentum space,  marked as LM for the main
sheet, LP for the Fermi pocket, and LS and LPS  for the others.
(b-f) show band structure (bottom panels) and corresponding momentum
distribution curves (MDCs) at the Fermi level (upper panels) along
five typical momentum cuts (cuts 1 to 5) as labeled in Fig. 1a.  To
see the weak features more clearly,  the original MDCs (thin grey
lines) in the upper panel  are expanded 10 times and plotted in the
same figures (thick black lines). Note that the signal of the Fermi
pocket LP is very weak; its intensity is over one order of magnitude
weaker than that of the main band LM.}
\end{figure}

%%\begin{figure*}[floatfix]
\begin{figure}[tbp]
\begin{center}
\includegraphics[width=1.0\columnwidth,angle=0]{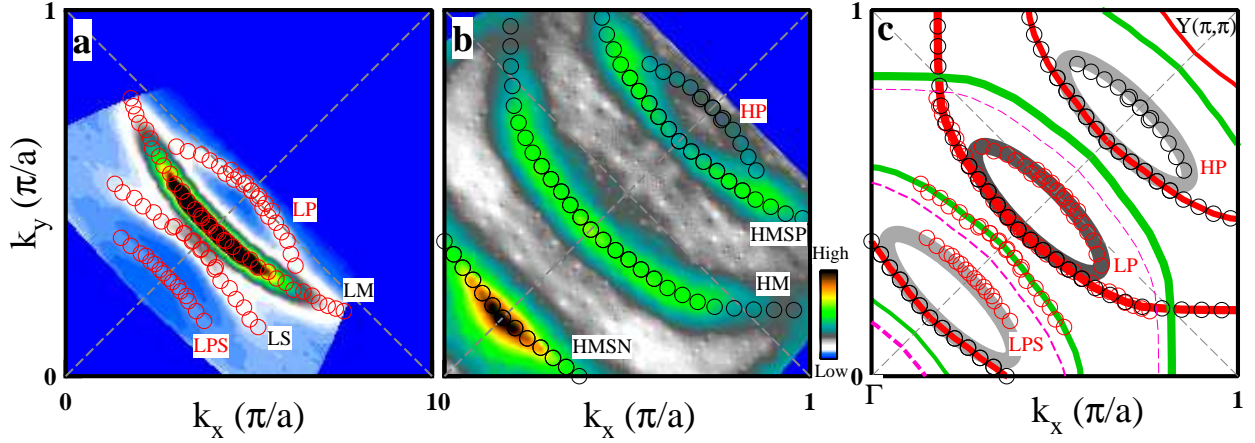}
\end{center}
\caption{Identification of the Fermi pocket in the photoemission
data. (a). Fermi surface measured in the La-Bi2201 UD18K sample by
the VUV laser. The four observed Fermi surface sheets in the covered
momentum space are quantitatively represented by red circles. (b).
Fermi surface measured in the La-Bi2201 UD18K sample by Helium
discharge lamp at a photon energy of 21.218 eV. It is obtained by
symmetrizing the original data with respect to the
(0,0)-($\pi$,$\pi$) line. Black circles represent quantitative
positions of the observed Fermi surface sheets.  (c). Summary of the
measured Fermi surface from VUV laser (red empty circles, as in Fig.
2a) and Helium discharge lamp (black empty circles, as in Fig. 2b).
Red lines represent the main Fermi surface (central thickest line)
and its corresponding umklapp bands (thinner lines on either side of
the main Fermi surface). The thickest green line represents the
shadow band of the main Fermi surface, and thinner green lines the
umklapp bands of the shadow band. The pink dashed lines represent
possible high order umklapp bands from the main band in the third
quadrant (see Supplementary and SFig. 2 for more details). It is
clear that the main Fermi surface measured from the VUV laser (LM)
shows a good agreement with that from the Helium discharge lamp
(HM). The LS band observed in VUV laser agrees well with the first
order umklapp band of the shadow band. The HMSP and HMSN bands
observed in Helium discharge lamp measurements obviously correspond
to the first order umklapp bands of the main band. The three
ellipses represent the position of the observed Fermi pockets. The
matrix element effect associated with photoemission process on the
relative spectral intensity between the main band and Fermi pocket
is an interesting issue to be further explored.}
%%\end{figure*}
\end{figure}

\begin{figure}[tbp]
\begin{center}
\includegraphics[width=1.0\columnwidth,angle=0]{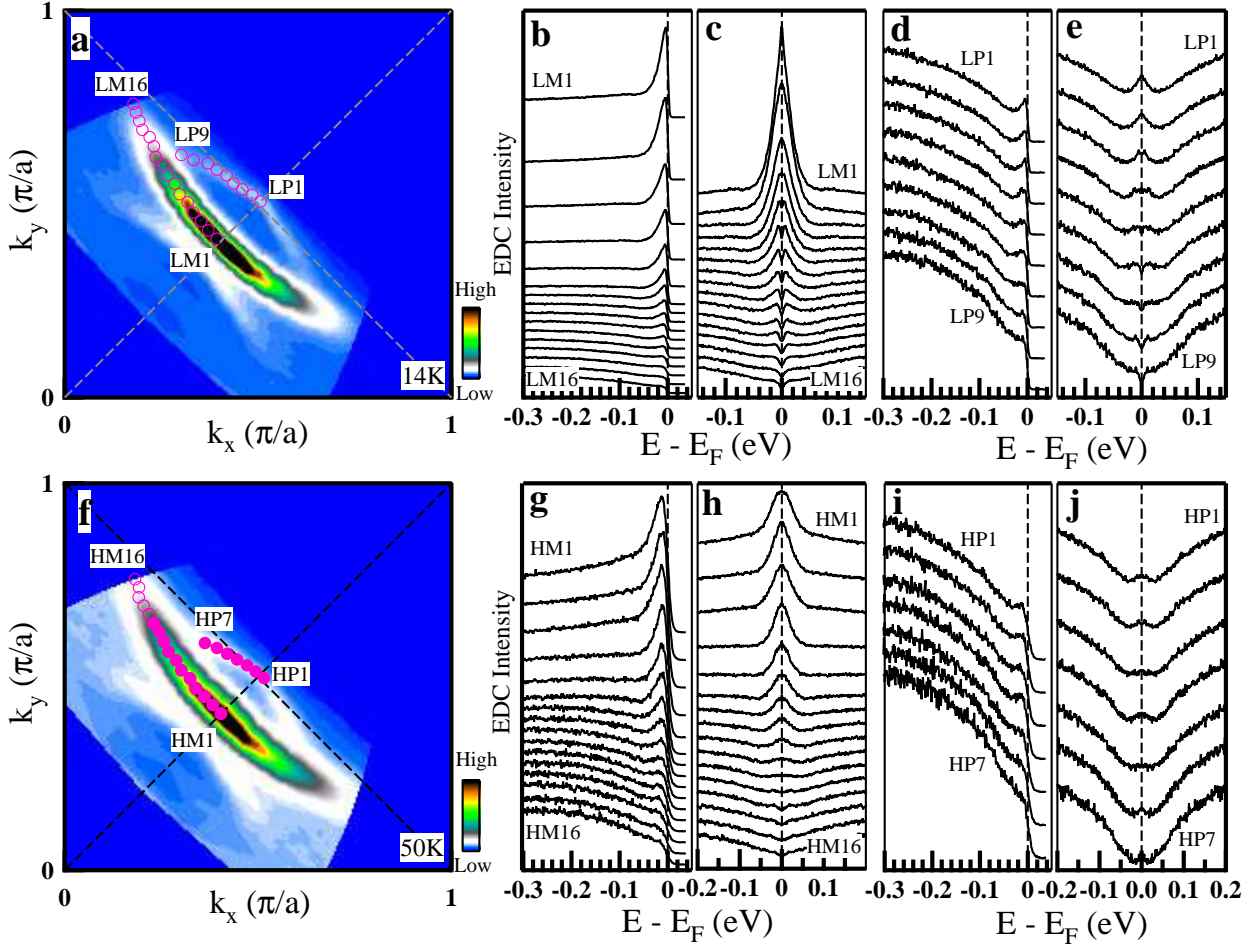}
\end{center}
\caption{Temperature dependence of the Fermi pocket. The top
(bottom) panels show Fermi surface and photoemission spectra for the
La-Bi2201 UD18K sample below T$_c$ (above T$_c$). (a). Fermi surface
mapping at 14 K. (b). Photoemission spectra (Energy Distribution
Curves, EDCs) along the main Fermi surface. Sharp peaks are observed
near the nodal region while they get weaker when moving to the
antinodal region. (c). The corresponding symmetrized EDCs. The
symmetrization procedure\cite{Norman} provides an intuitive way in
identifying an energy gap opening which is characterized by the
appearance of a dip near the Fermi level while zero gap corresponds
to a peak at the Fermi level. The gap size is determined by the EDC
peak position with respect to the Fermi level. It is clear to see
the gap opening and gap increase from the nodal to antinodal
regions. (d). EDCs along the back side of the Fermi pocket. (e). The
corresponding symmetrized EDCs. The gap opening and gap increase are
also clear from the nodal to antinodal regions along the pocket.
(f). Fermi surface mapping at 50 K. (g). EDCs along the main Fermi
surface. (h). The corresponding symmetrized EDCs. At 50 K, there is
a gapless region formed near the nodal region with its location
denoted in (f) by pink solid circles while the main Fermi surface
beyond this region remains gapped. (i). EDCs along the Fermi pocket.
(j). The corresponding symmetrized EDCs. At 50 K, the weak peak near
the Fermi level indicates gap closing along the Fermi pocket. }
\end{figure}

\begin{figure}[tbp]
\begin{center}
\includegraphics[width=1.0\columnwidth,angle=0]{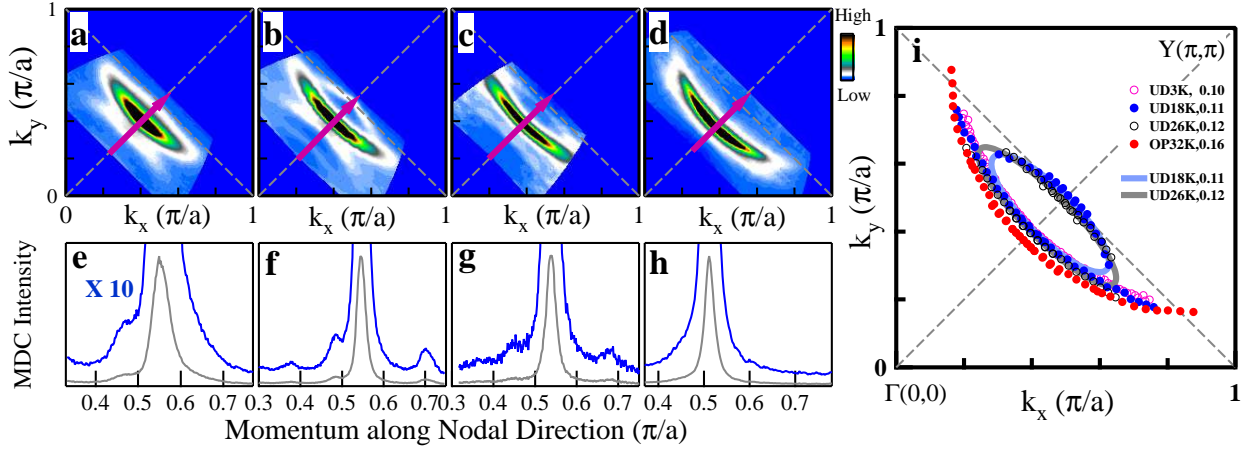}
\end{center}
\caption{Doping evolution of Fermi surface topology in La-Bi2201.
(a-d) show Fermi surface mapping of UD3K (underdoped, T$_c$$\sim$3
K), UD18K, UD26K (underdoped, T$_c$=26 K) and OP32K
(optimally-doped, T$_c$=32 K)\cite{JQMengdWave} samples,
respectively.  They are obtained by symmetrizing the original data
with respect to the (0,0)-($\pi$,$\pi$) line.  (e-h) show the
corresponding MDCs at the Fermi energy along the (0,0)-($\pi$,$\pi$)
nodal direction. The location of the momentum cuts is labeled in
(a-d) by purple lines with arrows. To show the weak features more
clearly, the original MDCs (thin grey lines in bottom panels in e-h)
are expanded by ten times and plotted in the same figures (blue
thick lines).  Note that we can not completely exclude the
possibility on the presence of Fermi pocket in the UD3K sample (Fig.
4a). It is noted that the MDC width of the sample (Fig. 4e) is much
wider than that of other dopings (Figs. 4f-h), leaving a possibility
that weak signal of a Fermi pocket may get buried in the broad
feature.  (i) Quantitative Fermi surface of La-Bi2201 at various
doping levels.  The two ``Fermi pockets" for UD18K and UD26K samples
are obtained by fitting both sides of the data points with
ellipses.}
\end{figure}

\end{document}